\documentstyle[preprint,aps]{revtex}
\tightenlines 

\begin{document}
\draft
\title{Nonequilibrium Fluctuations and Decoherence \\ in Nanomechanical Devices \\
Coupled to the Tunnel Junction. }
\author{Anatoly Yu. Smirnov}
\address{D-Wave Systems Inc., 320-1985 W. Broadway \\
Vancouver, British Columbia Canada V6J 4Y3}
\author{Lev G. Mourokh and Norman J.M. Horing}
\address{Department of Physics and Engineering Physics, \\
Stevens Institute of Technology, Hoboken, NJ 07030 }
\date{\today }
\maketitle

\begin{abstract}
{ We analyze the dynamics of a nanomechanical oscillator coupled to an electrical tunnel junction with an
arbitrary voltage applied to the junction and arbitrary temperature of electrons in leads. We obtain the explicit expressions for the fluctuations of oscillator position, its damping/decoherence rate, and the current through the structure. It is shown
that quantum heating of the oscillator results in nonlinearity of the current-voltage characteristics.  The effects of mechanical vacuum fluctuations are also discussed.

}
\end{abstract}

\pacs{85.85.+j, 73.63.-b, 03.65.Yz}

%\narrowtext

\section{Introduction}

The rapid development of nanotechnology in recent years has ushered in a new generation of quantum electronic
devices incorporating the mechanical degrees of freedom \cite{Roukes1,Craighead1}, so-called
nanoelectromechanical systems (NEMS). A particular example of such a device is the tunnel junction
having its transition matrix element modulated by a vibrational motion. This modulation can be achieved by the introduction of
small grains (shuttles) embedded in the elastic medium between the leads
\cite{Gorelik1,Gorelik2,Park1,Fedorets1,Armour1}, or by coupling the tunnel junction to a mechanical
oscillator (cantilever) \cite {Roukes2,Blencowe1,Mozyrsky1}. These two cases can be described by similar formulations but the oscillator masses are very different (for example, $m \sim 10^{-20}g$ for the
$C_{60}$ molecule used as a shuttle and $m \sim 10^{-12}g$ for the cantilever). Furthermore, in addition to nanomechanical applications, such systems have a direct relation to the fundamental quantum theory of
measurements \cite{Gurvitz1,Gurvitz2}. In this context, the tunnel junction can be consider as a device
measuring the position of the oscillator. As was shown in Ref.\cite{Mozyrsky1} for the case of a large
voltage applied to the junction, this measurement process induces decoherence and dephasing of the
oscillator, even at zero temperature. From another point of view, the nanomechanical oscillator can
serve as a measurement device for the electronic subsystem. A similar situation takes place in
magnetic resonance force microscopy \cite{Sidles1}, where the magnetic moment of electron or nuclear spins
in the sample (attached to the cantilever) is measured by means of optical interferometry of the
cantilever motion.

There are two manifestations of the effect of mechanical oscillations on system behavior and,
consequently, there are two characteristic scales. The first phenomenology occurs when the uncertainty of the oscillator
position is of the order of the tunneling length. This is always the case for the shuttle system,
but for the cantilever it only exists at high bias accompanied by oscillator heating. (In the present paper we simplify the model used in Refs.\cite{Gorelik1,Gorelik2,Park1}, such that our shuttle system has a permanent electrical connection to one lead, while the connection to the other lead is mediated by the tunnel
junction.)

The second class of phenomena is associated with resonance between the oscillator frequency and the bias voltage applied to the
tunnel junction. The characteristic frequencies of the oscillator are of order of
1GHz \cite{Roukes1,Roukes3} and this resonance occurs at sufficiently low
bias. It should be noted that the theoretical approach used by Mozyrsky and
Martin in Ref. \cite{Mozyrsky1} fails at low bias voltage and requires
improvment.

NEMS can be used as an ultrasensitive device for magnetic and biological applications and, therefore,
its own fluctuation level - determining the signal/noise ratio - is of crucial importance. Moreover, quantum measurement procedures also require a high level of sensitivity. The mechanical oscillator can be considered as an open
quantum system interacting with two electron reservoirs (leads) serving as an
effective heat bath. In the present paper we employ the general theory of open quantum systems
developed in Refs. \cite {Efremov1,Efremov2,Smirnov1} to determine the nonequilibrium fluctuations of the
mechanical oscillator for arbitrary bias voltage and temperature. Current noise created by the
junction is assumed to be the main mechanism of decoherence of the oscillator.

We derive general formulas and employ them in the two cases described above, showing that at low temperature the
bias applied to the tunnel junction leads to oscillator heating at voltages larger than the critical
value associated with the characteristic frequency of the oscillator. With increasing temperature this
resonant picture is smoothed and, moreover, for higher temperature the fluctuation level approaches that of an uncoupled
oscillator. Furthermore, we show that in the shuttle case, a nonlinearity of
the oscillator-junction interaction  makes a pronounced contribution to the level of mechanical vacuum
fluctuations even for zero bias and weak coupling between the above-mentioned subsystems.

\section{General Formalism}

The Hamiltonian of the system {\it tunnel junction and mechanical oscillator} is
given by
\begin{equation}
H = H_L + H_R + H_{tun} + H_0.
\end{equation}
Here,
\begin{equation}
H_{\alpha } = \sum_k E_{k\alpha } c_{k\alpha}^+c_{k\alpha}
\end{equation}
is the Hamiltonian of the left, right leads $(\alpha = L,R$, respectively)
\begin{equation}
H_0 = {\frac{p^2 }{2m}} + {\frac{m\omega_0^2 x^2 }{2}}
\end{equation}
is the Hamiltonian of the mechanical oscillator with a mass $m$ and the
resonant frequency $\omega_0,$, and
\begin{equation}
H_{tun} = - \sum_{kq}(T_{kq}c_{kL}^+c_{qR} + h.c.) e^{-x/\lambda }
\end{equation}
is a tunneling term depending on a position of the oscillator, where $\lambda
$ is the characteristic tunneling length and $h.c.$ is the Hermitian
conjugate. The electron gas in the leads plays the role of a heat bath for the
oscillator and their nonlinear interaction given in Eq.(4) can be written in the form
\begin{equation}
H_{tun} = - Q e^{-x/\lambda }
\end{equation}
where $Q(t)$ is interpreted as an effective heat bath variable given by
\begin{equation}
Q(t) = \sum_{kq}(T_{kq}c_{kL}^+c_{qR} + h.c.).
\end{equation}
The heat bath thus defined is characterized by a response function $\varphi (t,t_1)$ and a symmetrized correlation function $M(t,t_1)$ of the unperturbed variables
$Q^{(0)}(t)$ taken in the absence of interaction ($\hbar =1, k_B = 1$),
as
\begin{eqnarray}
\varphi(t,t_1) = i\langle [Q^{(0)}(t),Q^{(0)}(t_1)]_-\rangle \theta(t-t_1),
\nonumber \\
M(t,t_1) = {1\over 2}\langle [Q^{(0)}(t),Q^{(0)}(t_1)]_+\rangle ,
\end{eqnarray}
where $\theta(\tau )$ is the unit Heaviside step function and $%
[...,...]_{+}$  and $ [...,...]_{-}$ are the anticommutator and the commutator,
respectively. The angle brackets refer to averaging over the equilibrium states
of both the left and right leads at the same information. The chemical potentials of the leads can be
different with $\mu_R - \mu_L = eV,$ where $V$ is a voltage applied to the
tunnel junction. For weak coupling between the mechanical oscillator and the
electronic bath (weak tunneling) the action of the oscillator on the
dissipative environment is described by the formula \cite{Efremov1}
\begin{equation}
Q(t) = Q^{(0)}(t) + \int dt_1 \varphi(t,t_1) e^{-x(t_1)/\lambda }.
\end{equation}
In turn, the effect of the dissipative environment on the mechanical oscillator is determined by substituting Eq.(8) into the Heisenberg equation of motion
for the position operator of the oscillator given by
\begin{equation}
\ddot{x} + \omega_0^2 x = - {\frac{1}{m\lambda }} Q(t)e^{-x/\lambda}.
\end{equation}
To eliminate the unperturbed heat bath variables $Q^{(0)}(t)$ from Eq.(9), we
apply the quantum Furutsu-Novikov theorem \cite{Efremov1}
\begin{equation}
\langle Q^{(0)}(t)e^{-x(t)/\lambda}\rangle = \int dt_1 \langle
Q^{(0)}(t),Q^{(0)}(t_1) \rangle \langle {\frac{\delta }{\delta Q^{(0)}(t_1)}}
e^{-x(t)/\lambda}\rangle,
\end{equation}
where $\delta /\delta Q^{(0)}(t_1)$ is the functional derivative with
respect to the free heat bath variable $Q^{(0)}(t_1)$. The functional
derivative of an arbitrary operator $A(t)$ of the dynamical system is
proportional to the commutator \cite{Efremov1} in the form
\begin{equation}
{\frac{\delta A(t) }{\delta Q^{(0)}(t_1)}} = {\frac{i}{\hbar }}\left[
A(t),e^{-x(t_1)/\lambda}\right]_-\theta(t-t_1).
\end{equation}
As the result, we obtain the non-Markovian stochastic equation of motion for
the position operator of the oscillator as given by
\begin{eqnarray}
\ddot{x} + \omega_0^2 x = \xi (t)  \nonumber \\ -
{\frac{1 }{\lambda m}}\int dt_1 \left( \tilde{M}(t,t_1) i\left[%
e^{-x(t)/\lambda},e^{-x(t_1)/\lambda}\right]_- + \varphi(t,t_1){\frac{1}{2}} %
\left[e^{-x(t)/\lambda},e^{-x(t_1)/\lambda}\right]_+\right),
\end{eqnarray}
where $\tilde{M}(t,t_1) = M(t,t_1)\theta(t-t_1)$ is the causal correlation
function of the heat bath and $\xi(t)$ is the fluctuation source,
\begin{equation}
\xi(t) = -{\frac{1}{m\lambda }}\left( Q^{(0)}(t)e^{-x(t)/\lambda} - \int
dt_1 \langle Q^{(0)}(t),Q^{(0)}(t_1)\rangle i\left[e^{-x(t)/%
\lambda},e^{-x(t_1)/\lambda}\right]_- \theta(t-t_1) \right) ,
\end{equation}
having zero mean value, $\langle \xi \rangle = 0$, according to Eq.(10). The Langevin-like equation, Eq.(12), as well as the whole method of Refs. \cite{Efremov1,Efremov2,Smirnov1}, takes into account a nonlinearity
of coupling between the subsystems and, in addition, incorporates the nonlocal character of heat bath fluctuations. In these respect this treatment goes
beyond the well-known Caldeira-Leggett approach \cite{Caldeira1}.

In the case of weak coupling, the correlation function of fluctuation
sources is given by
\begin{eqnarray}
\langle {\frac{1}{2}} \left[ \xi(t), \xi(t^{\prime})\right]_+\rangle = {%
\frac{1 }{4m^2\lambda^2}} \{\langle
[Q^{(0)}(t),Q^{(0)}(t^{\prime})]_+\rangle \langle \left[e^{-x(t)/%
\lambda},e^{-x(t^{\prime})/\lambda}\right]_+\rangle  \nonumber \\ +
\langle [Q^{(0)}(t),Q^{(0)}(t^{\prime})]_- \rangle \langle \left[%
e^{-x(t)/\lambda},e^{-x(t^{\prime})/\lambda}\right]_-\rangle \}.
\end{eqnarray}
It should be mentioned that the fluctuations of electronic variables $%
\{Q^{(0)}(t)\}$ are non-Gaussian. However, as indicated above, Eqs.(8),(10),(14) are valid for weak coupling between the dynamical system (mechanical oscillator) and the
dissipative environment (electrons in the leads). In this case we
can calculate the (anti)commutators in Eqs.(12),(14) using a free evolution
approximation
\begin{equation}
x(t) = x(t_1) \cos\omega_0(t-t_1) + {\frac{p(t_1) }{m\omega_0}}
\sin\omega_0(t-t_1),
\end{equation}
and employing the Baker-Hausdorff theorem, we obtain
\begin{eqnarray}
{\frac{1}{2}}\left[e^{-x(t)/\lambda},e^{-x(t_1)/\lambda}\right]_+ =
\cos(\nu_0 \sin\omega_0\tau) \exp\left(- {\frac{x(t) + x(t_1)}{\lambda }}%
\right),  \nonumber \\
i\left[e^{-x(t)/\lambda},e^{-x(t_1)/\lambda}\right]_- = 2\sin(\nu_0
\sin\omega_0\tau) \exp\left(- {\frac{x(t) + x(t_1)}{\lambda }}\right).
\end{eqnarray}
Here,
\begin{equation}
\nu_0 = {\frac{\hbar }{2 m\omega_0\lambda^2}}
\end{equation}
is the square of the ratio of the unperturbed oscillator position uncertainty at zero temperature and
the tunneling
length $\lambda.$ The operator $x(t)$ can be written as a sum of the mean
and fluctuating parts, $x(t) = \bar{x}(t) + \tilde{x}(t).$ Even if the
averaged position of the oscillator is small compared to the tunneling
length, $\bar{x}(t) \ll \lambda$, the fluctuating amplitude can be of the order
of the tunneling length causing nonlinear effects. To examine this
nonlinearity, we assume that oscillator position fluctuations are
approximately described by Gaussian statistics with a dispersion $\langle
\tilde{x}^2 \rangle = \langle \tilde{x}(t) \tilde{x}(t)\rangle.$ In this
approximation, we obtain the exponent involved in Eq.(16) as
\begin{equation}
\exp\left(- {\frac{x(t) + x(t_1)}{\lambda }}\right) = \left[ 1 - {\frac{%
\tilde{x}(t) + \tilde{x}(t_1)}{\lambda }}- {\frac{\bar{x}(t) + \bar{x}(t_1)}{%
\lambda }} \right] \exp\left({\frac{\langle \tilde{x}^2 \rangle + \langle
(1/2)[\tilde{x}(t),\tilde{x}(t_1)]_+ \rangle }{\lambda^2}}\right).
\end{equation}
Eqs.(16) and (18) provide the expressions required on the right side of the non-Markovian stochastic
equation, Eq.(12). The dispersion $\langle \tilde{x}^2 \rangle $ and the
correlator $\langle (1/2)[\tilde{x}(t),\tilde{x}(t_1)]_+\rangle $ have to be
determined self-consistently. To accomplish this, we suppose that the operator $%
\tilde{x}$ of the oscillator position obeys the equation
\begin{equation}
\ddot{\tilde{x}} + \gamma \dot{\tilde{x}} + \omega_0^2 \tilde{x} = \xi,
\end{equation}
where the damping rate $\gamma$  can be calculated from the collision term on the right of Eq.(12). Using Eq.(19), we obtain
correlator and the dispersion of the oscillator position as
\begin{eqnarray}
\langle \tilde{x}^2 \rangle = {\frac{K(\omega_0) }{2 \omega_0 \gamma}},
\nonumber \\
\langle {\frac{1}{2}}[\tilde{x}(t),\tilde{x}(t_1)]_+\rangle = \langle \tilde{%
x}^2 \rangle e^{-\gamma |t-t_1| /2} \cos \omega_0(t-t_1).
\end{eqnarray}
Here,
\begin{equation}
K(\omega ) = \int d\tau e^{i\omega \tau }\langle {\frac{1}{2}} \left[
\xi(\tau), \xi (0)\right]_+\rangle
\end{equation}
is the spectral function of the fluctuation forces, Eq.(14). Assuming that the averaged deviation of the
oscillator position is much less than the tunneling length $\lambda $, $\bar{x} \ll \lambda,$ (the
root-mean-square amplitude of mechanical oscillations can be of the order of $\lambda$), we obtain a
simplified equation for the total coordinate $x = \bar{x} + \tilde{x}$ as
\begin{equation}
\ddot{x} + \omega_0^2 x + {\frac{1}{m\lambda^2}}\int dt_1 L(t-t_1)[x(t) + x(t_1)] = \xi (t) -
{\frac{1}{m}} F_0,
\end{equation}
where the collision kernel $L(\tau )$ is given by the expression
\begin{equation}
L(\tau ) = \left[ 2\tilde{M}(\tau )\sin(\nu_0 \sin\omega_0\tau ) + \varphi(\tau ) \cos(\nu_0
\sin\omega_0\tau )\right] \exp[\nu_c(1 +\cos\omega_0\tau )]
\end{equation}
with the parameter
\begin{equation}
\nu_c = {\frac{\langle \tilde{x}^2 \rangle }{\lambda^2}},
\end{equation}
describing nonequilibrium fluctuations of the oscillator position. The deterministic force $F_0$ on the right side of Eq.(22), which arises from interaction of the oscillator with the tunnel junction, is given by $F_0 = (1/\lambda)
\Lambda(\omega = 0), $ where $\Lambda(\omega )$ is the Fourier transform of the kernel $L(\tau)$,
Eq.(23). It follows from Eq.(22), that coupling to the tunnel junction also results in a shift of
the oscillator frequency, $\omega_0^2 \rightarrow \omega_0^2 - (1/m\lambda^2) \Lambda(\omega = 0),$
jointly with damping and decoherence described by $Im\Lambda(\omega) \equiv\Lambda^{\prime\prime}(\omega ).
$

The relaxation rate of mechanical oscillations due to coupling to the
tunnel junction is given by
\begin{equation}
\gamma = {\frac{\hbar }{m\lambda^2}}{\frac{\Lambda^{\prime\prime}(\omega_0)}{%
\omega_0}},
\end{equation}
where
\begin{eqnarray}
\Lambda^{\prime\prime}(\omega) = {\frac{1}{2}}e^{\nu_c}
\sum_{l=-\infty}^{l=+\infty}I_l(\nu_c)J_0(\nu_0) \{\chi^{\prime\prime}(\omega - l\omega_0) +
\chi^{\prime\prime}(\omega +
l\omega_0)\}   \nonumber \\ +
{\frac{1}{2}}e^{\nu_c}
\sum_{l=-\infty}^{l=+\infty}\sum_{n=1}^{n=\infty}I_l(\nu_c)J_{2n}(\nu_0)
\{\chi^{\prime\prime}(\omega +2n\omega_0 - l\omega_0) +
\chi^{\prime\prime}(\omega - 2n\omega_0 + l\omega_0)  \nonumber \\ +
\chi^{\prime\prime}(\omega + 2n\omega_0 + l\omega_0) +
\chi^{\prime\prime}(\omega - 2n\omega_0 - l\omega_0)\}   \nonumber \\ +
{\frac{ 1}{2}}e^{\nu_c}
\sum_{l=-\infty}^{l=+\infty}\sum_{n=0}^{n=\infty}I_l(\nu_c)J_{2n+1}(\nu_0)
  \nonumber \\ \times 
\{S[\omega - (2n+1)\omega_0-l\omega_0] + S[\omega -
(2n+1)\omega_0+l\omega_0]  \nonumber \\ -
S[\omega + (2n+1)\omega_0-l\omega_0] - S[\omega + (2n+1)\omega_0 +
l\omega_0]\}.
\end{eqnarray}
Here, $J_n(\nu)$ is the Bessel function of n-th order, and $I_l(\nu)$ is the modified Bessel function
of l-th order. The spectral function $S(\omega )$ and the imaginary part of the susceptibility
$\chi^{\prime\prime}(\omega )$ are the Fourier transforms of the correlator and the response function
of the heat bath, Eq.(7), respectively,
\begin{eqnarray}
S(\omega ) = \int d\tau e^{i\omega \tau } M(\tau),  \nonumber \\
\chi^{\prime\prime}(\omega ) = \int d\tau \sin(\omega \tau ) \varphi (\tau ).
\end{eqnarray}
For a harmonic oscillator, the inverse relaxation rate $\gamma^{-1} $ given by Eq.(25) also represents a
characteristic time scale for the coherence decay, $\tau_d = \gamma^{-1}.$

The parameter $\nu_c$ of Eq.(24) describing the level of mechanical fluctuations can be found from the
self-consistent equation
\begin{equation}
\nu_c = {\frac{K(\omega_0) }{2 \omega_0^2 \lambda^2 \gamma}},
\end{equation}
where $K(\omega ) $ is given by
\begin{eqnarray}
K(\omega ) = {\frac{\hbar^2}{2 m^2 \lambda^2 }}e^{\nu_c}
\sum_{l=-\infty}^{l=+\infty}I_l(\nu_c)J_0(\nu_0) \{S(\omega - l\omega_0) +
S(\omega + l\omega_0)\}  \nonumber \\ +
{\frac{1}{2}}e^{\nu_c}
\sum_{l=-\infty}^{l=+\infty}\sum_{n=1}^{n=\infty}I_l(\nu_c)J_{2n}(\nu_0)
\{S(\omega +2n\omega_0 - l\omega_0) + S(\omega - 2n\omega_0 + l\omega_0) 
\nonumber \\ +
S(\omega + 2n\omega_0 + l\omega_0) + S(\omega - 2n\omega_0 - l\omega_0)\} 
\nonumber \\ -
{\frac{1}{2}}e^{\nu_c}
\sum_{l=-\infty}^{l=+\infty}\sum_{n=0}^{n=\infty}I_l(\nu_c)J_{2n+1}(\nu_0)
\{\chi^{\prime\prime}[\omega +(2n+1)\omega_0 - l\omega_0] -
\chi^{\prime\prime}[\omega - (2n+1)\omega_0 - l\omega_0]  \nonumber \\ +
\chi^{\prime\prime}[\omega + (2n+1)\omega_0 + l\omega_0] -
\chi^{\prime\prime}[\omega - (2n+1)\omega_0 + l\omega_0]\} .
\end{eqnarray}

\section{Reservoir correlation functions and tunnel current}

To determine the oscillator damping rate $\gamma $ of Eq.(25) and the level of
fluctuations $\nu _{c}$ of Eq.(24), we need to determine the spectral function $%
S(\omega )$ and the susceptibility $\chi ^{\prime \prime }(\omega )$ of the
electronic heat bath, Eq.(27), for the case with {\it no} interaction with
the oscillator. Assuming that the variables $\{c_{kL}^{(0)},c_{qR}^{(0)},..\}
$ of the free electron gases in the leads obey the Wick theorem (with
uncorrelated electron systems in the different leads), we can express the
correlator of unperturbed heat bath variables in the form
\begin{eqnarray}
\langle Q^{(0)}(t)Q^{(0)}(t_{1})\rangle  &=&\sum_{kq}|T_{kq}|^{2}\{\langle
c_{kL}^{(0)+}(t)c_{kL}^{(0)}(t_{1})\rangle \langle
c_{qR}^{(0)}(t)c_{qR}^{(0)+}(t_{1})\rangle  \nonumber \\ +
&&\langle c_{kL}^{(0)}(t)c_{kL}^{(0)+}(t_{1})\rangle \langle
c_{qR}^{(0)+}(t)c_{qR}^{(0)}(t_{1})\rangle \}.
\end{eqnarray}
Introducing retarded, advanced and ''lesser'' Green's functions of
electrons in the leads as
\begin{eqnarray}
g_{k\alpha }^{r}(t,t_{1}) &=&\langle (-i)[c_{k\alpha }^{(0)}(t),c_{k\alpha
}^{(0)+}(t_{1})]_{+}\rangle \theta (t-t_{1})=-ie^{-iE_{k\alpha
}(t-t_{1})}\theta (t-t_{1}),  \nonumber \\
g_{k\alpha }^{a}(t,t_{1}) &=&\langle i[c_{k\alpha }^{(0)}(t),c_{k\alpha
}^{(0)+}(t_{1})]_{+}\rangle \theta (t_{1}-t)=ie^{-iE_{k\alpha
}(t_{1}-t)}\theta (t_{1}-t),  \nonumber \\
g_{k\alpha }^{<}(t,t_{1}) &=&i\langle c_{k\alpha }^{(0)}(t),c_{k\alpha
}^{(0)+}(t_{1})\rangle \theta (t-t_{1})=if_{\alpha }(E_{k\alpha
})e^{-iE_{k\alpha }(t-t_{1})},
\end{eqnarray}
where
\begin{equation}
f_{\alpha }(E)=f(E-\mu _{\alpha })=\left[ \exp \left( {\frac{E-\mu _{\alpha }%
}{T}}\right) +1\right] ^{-1}
\end{equation}
is the Fermi distribution in the $\alpha -$lead ($\alpha =L,R$) with the
chemical potential $\mu _{\alpha }$ and temperature $T$, we obtain
\begin{eqnarray}
M(t,t_{1}) &=&{\frac{1}{2}}\sum_{kq}|T_{kq}|^{2}%
\{g_{qR}^{<}(t,t_{1})[g_{kL}^{r}(t_{1},t)-g_{kL}^{a}(t_{1},t)] \nonumber \\ 
&&+\lbrack
g_{qR}^{r}(t,t_{1})-g_{qR}^{a}(t,t_{1})]g_{kL}^{<}(t_{1},t)+g_{kL}^{<}(t,t_{1})
[g_{qR}^{r}(t_{1},t)-g_{qR}^{a}(t_{1},t)]
\nonumber \\ 
&&+\lbrack
g_{kL}^{r}(t,t_{1})-g_{kL}^{a}(t,t_{1})]g_{qR}^{<}(t_{1},t)+2g_{qR}^{<}(t,t_{1})g_{kL}^{<}
(t_{1},t)+2g_{kL}^{<}(t,t_{1})g_{qR}^{<}(t_{1},t)\},
\nonumber \\
\varphi (t,t_{1}) &=&i\theta
(t-t_{1})\sum_{kq}|T_{kq}|^{2}%
\{[g_{qR}^{r}(t,t_{1})-g_{qR}^{a}(t,t_{1})]g_{kL}^{<}(t_{1},t)-g_{qR}^{<}(t,t_{1})[g_{kL}^{r}(t_{1},t)-
g_{kL}^{a}(t_{1},t)]
\nonumber \\ 
&&+\lbrack
g_{kL}^{r}(t,t_{1})-g_{kL}^{a}(t,t_{1})]g_{qR}^{<}(t_{1},t)-g_{kL}^{<}(t,t_{1})[g_{qR}^{r}(t_{1},t)-
g_{qR}^{a}(t_{1},t)]\}.
\end{eqnarray}
As a result, the formulas for the spectrum and the imaginary part of the
susceptibility of the dissipative environment are given by
\begin{eqnarray}
S(\omega ) &=&\pi \sum_{kq}|T_{kq}|^{2}[\delta (\omega
-E_{kL}+E_{qR})+\delta (\omega +E_{kL}-E_{qR})]  \nonumber \\
&&\times \lbrack f_{R}(E_{qR})+f_{L}(E_{kL})-2f_{L}(E_{kL})f_{R}(E_{qR})],
\nonumber \\
\chi ^{\prime \prime }(\omega ) &=&\pi \sum_{kq}|T_{kq}|^{2}[\delta (\omega
+E_{kL}-E_{qR})-\delta (\omega -E_{kL}+E_{qR})][f_{L}(E_{kL})-f_{R}(E_{qR})].
\end{eqnarray}
It should be mentioned that when the leads are in the full thermodynamic equilibrium state,
$(\mu _{L}=\mu _{R}),$ the spectral function $S(\omega )$ and the
function $\chi ^{\prime \prime }(\omega )$ are related by means of the
fluctuation-dissipation theorem $S(\omega )=\chi ^{\prime \prime }(\omega
)\coth (\omega /2T).$

The electric current through the junction is defined as $I = I_L = -I_R,$
where $I_{\alpha} =  e \langle \dot{N}_{\alpha}\rangle $ with $N_{\alpha } =
\sum_k c^+_{k\alpha }c_{k\alpha}$ being an electron  number in the $\alpha-$%
lead. It follows from the equations of motion for the electron operators of the leads,
\begin{eqnarray}
i \dot{c}_{pL} = E_{pL} c_{pL} - \sum_{q^{\prime}} T_{pq^{\prime}}
c_{q^{\prime}R} e^{-x/\lambda },  \nonumber \\
i \dot{c}_{qR} = E_{qR} c_{qR} - \sum_{k^{\prime}} T_{p^{\prime}q}
c_{p^{\prime}L} e^{-x/\lambda },
\end{eqnarray}
that the electric current depends on the oscillator position $x$ as
\begin{equation}
I = -i \sum_{pq}\langle \{T_{pq}^*c^+_{qR}c_{pL} - h.c.\}
e^{-x/\lambda}\rangle .
\end{equation}
In the case of weak tunneling between leads, the electron operators of the leads
can be represented as
\begin{eqnarray}
c_{pL}(t) = c_{pL}^{(0)}(t) - \sum_{q^{\prime}}T_{pq^{\prime}}\int dt_1
g^r_{pL}(t,t_1)c_{q^{\prime}R}^{(0)}(t_1) e^{-x(t_1)/\lambda},  \nonumber \\
c_{qR}(t) = c_{qR}^{(0)}(t) - \sum_{p^{\prime}}T_{p^{\prime}q}^*\int dt_1
g^r_{qR}(t,t_1)c_{p^{\prime}L}^{(0)}(t_1) e^{-x(t_1)/\lambda}.
\end{eqnarray}
Therefore, the electric current is given by
\begin{eqnarray}
I = \sum_{pq}|T_{pq}|^2 \int dt_1 \{ g^r_{pL}(t,t_1)g^<_{qR}(t_1,t) 
\nonumber \\
-g^r_{qR}(t,t_1)g^<_{pL}(t_1,t) \} \langle {\frac{1}{2}}\left[
e^{-x(t)/\lambda }, e^{-x(t_1)/\lambda }\right]_+\rangle + h.c. .
\end{eqnarray}
Substituting the expressions for the anticommutator, Eq.(16), we obtain
\begin{eqnarray}
I = {\frac{\pi}{2}} e^{\nu_c}
\sum_{l=-\infty}^{l=+\infty}\sum_{n=-\infty}^{n=+\infty}I_l(\nu_c)J_{2n}(%
\nu_0)\sum_{pq}|T_{pq}|^2 [f_R(E_{qR}) - f_L(E_{kL})] \nonumber \\
\times  \{ \delta[E_{pL} - E_{qR} + (l-2n)\omega_0] + \delta[E_{pL} - E_{qR} -
(l-2n)\omega_0]  \nonumber \\
+\delta[E_{pL} - E_{qR} + (l+2n)\omega_0] + \delta[E_{pL} - E_{qR} -
(l+2n)\omega_0] \}.
\end{eqnarray}

We can determine the spectral function $S(\omega )$ and the imaginary
part of the susceptibility $\chi^{\prime\prime}(\omega )$ of the
electron reservoir, Eq.(34), by introducing the densities-of-states $D_{\alpha}(E)$ of the
leads and replacing sums over $k,q,..$ by integrations over the
corresponding energies as
\[
\sum_{kq} (...) \rightarrow \int dE_L dE_R D_L(E_L)D_R(E_R)(...).
\]
We assume that the tunneling elements $|T_{kq}|^2$ do not depend on
energies, $|T_{kq}|^2 = |T_0|^2$, and the densities-of-states near the Fermi
surface in the leads are also energy-independent, $D_{\alpha }(E)\simeq
D_{\alpha}(\mu ) = D_{\alpha }$ with $\mu_R = \mu +eV/2, \mu_L = \mu -eV/2, \alpha
= L,R.$ Furthermore, we assume that all energy parameters of our problem (%
$eV,T,\omega_0, ..$) are much less than the basic chemical potential $\mu$
of the electron gas in the leads. In this case, we obtain the imaginary part of the
susceptibility as
\begin{equation}
\chi^{\prime\prime}(\omega ) = \pi D_L D_R |T_0|^2\int_{-\infty}^{+\infty}
dE [f(E-\omega ) - f(E+\omega)],
\end{equation}
whereas the spectral function $S(\omega) $ has the form
\begin{eqnarray}
S(\omega) = \pi D_L D_R |T_0|^2\int_{-\infty}^{+\infty} dE \{ [1 -
f(E-\mu_L)] [f(E-\mu_R +\omega) + f(E-\mu_R -\omega) ] +  \nonumber \\
f(E-\mu_L)[2 - f(E-\mu_R +\omega) - f(E-\mu_R -\omega)]\}.
\end{eqnarray}
As a result, the electrons in the leads represent an Ohmic heat bath with
respect to the mechanical oscillations of the cantilever with the imaginary part of susceptibility given by
\begin{equation}
\chi^{\prime\prime}(\omega ) = \alpha \omega,
\end{equation}
where $ \alpha =  2\pi D_L D_R |T_0|^2 $, and the frequency-dependent spectral function has the form
\begin{equation}
S(\omega ) = {\alpha \over 2} \left[ (\omega - eV) \coth \left({\frac{%
\omega - eV }{2T}}\right) +(\omega + eV) \coth \left({\frac{\omega + eV }{2T}%
}\right)\right].
\end{equation}
At zero temperature $S(\omega )$ is given by
\begin{equation}
S(\omega ) = \alpha [ \omega \theta(\omega - eV) + eV \theta(eV - \omega )],
\end{equation}
where $\theta(\omega )$ is the Heaviside step function.

For the electric current, we finally obtain
\begin{equation}
I = G(V)V = \alpha e^{2\nu_c} eV,
\end{equation}
where $G(V)= e\alpha \exp(2\nu_c)$ is the nonlinear conductance of the tunnel junction, which depends on the fluctuation level of the mechanical oscillator.

\section{Resuts and discussion}

Although the general formalism developed in the previous sections makes it possible to perform comprehensive analyses of the rich phenomenology occurring in a wide variety of nanoelectromechanical systems, we have restricted ourselves in the
present paper to the examination of nonequilibrium fluctuations of a mechanical
oscillator coupled to a biased tunnel junction, along with the study of quantum heating effects
on the decoherence rate (Eq.(25)) of the oscillator  and on the current-voltage characteristics (Eq.(45)) of the
junction.

To start, we demonstrate analytical results related to the case of  weak nonlinearity of the
cantilever-junction coupling, Eq.(5), when $\nu_0 \ll 1$. In this case, the contribution of vacuum
fluctuations to the collision term $\Lambda^{\prime\prime}(\omega)$, Eq.(26), and to the correlator $K(\omega )$, Eq.(29), can be neglected. Therefore, the dissipative kernel is proportional to
$\omega$, $\Lambda^{\prime\prime}(\omega) = \alpha \exp(2\nu_c)\omega.$ This corresponds to Eq.(19) for the oscillator position with the damping rate $\gamma $, Eq.(25), and the spectrum of
the fluctuation forces $K(\omega )$, Eq.(29), given by 
\begin{eqnarray}
\gamma = \alpha {\hbar \over m\lambda^2} e^{2\nu_c}, \nonumber\\
K(\omega ) = {\hbar^2 \over 2 m^2\lambda^2} e^{\nu_c}\sum_{l=-\infty}^{l=+\infty}I_l(\nu_c)
 [S(\omega - l\omega_0) + S(\omega + l\omega_0)].
\end{eqnarray}
In contrast to the approach of Ref. \cite{Mozyrsky1}, the fluctuations of the random forces $\xi$ involved in the Langevin
equation, Eq.(19), are not white noise because of the frequency dispersion of the
correlator $K(\omega )$, Eq.(46). Even for the case of weak heating, when the root-mean-square amplitude of
the cantilever fluctuations is much less than the tunneling length $\lambda$, $\sqrt{\langle \tilde{x}^2
\rangle} \ll \lambda,$ and
 $\nu_c \ll 1,$
the spectrum $K(\omega )$ of the fluctuation sources,
\begin{equation}
K(\omega ) = \alpha {\hbar^2 \over  2m^2\lambda^2} e^{2\nu_c} \left[ (\omega - eV) \coth \left({\frac{%
\omega - eV }{2T}}\right) +(\omega + eV) \coth \left({\frac{\omega + eV }{2T}%
}\right)\right],
\end{equation}
demonstrates a non-white character, especially at zero temperatures of the electron reservoirs.

On the basis of the above-mentioned approximations and with Eqs. (28), (46), (47) we have derived the dispersion of the cantilever fluctuations $\langle \tilde{x}^2 \rangle
= \lambda^2 \nu_c$ as a function of voltage $V$ applied to the junction and of temperature $T$ of
the electron reservoirs, as
\begin{equation}
\langle \tilde{x}^2 \rangle  = {\hbar \over 4 m\omega_0}\left[ {\omega_0 - eV\over \omega_0} \coth
\left({\omega_0 - eV \over 2T}\right) +{\omega_0 + eV\over \omega_0} \coth \left({\frac{\omega_0 + eV
}{2T} }\right)\right].
\end{equation}
At low temperatures, $T \ll \omega_0 \pm eV,$ and at low voltages, $eV < \omega_0, $ when the current
in the tunnel junction can not stimulate an excitation of the mechanical system, the dispersion of
the oscillator fluctuations remains on the vacuum level, $\langle \tilde{x}^2 \rangle  = (\hbar /2
m\omega_0).$ However, at the higher voltage, $eV > \omega_0,$ the level of
fluctuations increases linearly with the voltage, as $\langle \tilde{x}^2 \rangle  = (\hbar /2
m\omega_0)(eV/\omega_0)$, and we find the dimensionless parameter $\nu_c$ as $\nu_c = \nu_0 (eV/\omega_0). $ This linear increase of the fluctuation level may be interpreted as an increase of an effective temperature of the oscillator: $T_{eff} = eV/2,$
where  $T_{eff}$  is proportional to the dispersion of mechanical fluctuations, $ \langle \tilde{x}^2
\rangle = T_{eff} /m\omega_0^2. $ This confirms the result obtained by Mozyrsky and Martin
\cite{Mozyrsky1} for the Caldeira-Leggett model.

The heating process affects also the bias dependence of the damping rate $\gamma $, Eq.(46), and
the behavior of the conductance $G(V)$, Eq.(45). At low voltages, $eV < \hbar \omega_0, $ both of
these characteristics do not depend on
$V, [G = e\alpha (1 + 2\nu_0), \gamma = (\hbar \alpha /m\lambda^2)(1 +
2\nu_0)], $ whereas above the threshold of the excitation of cantilever oscillations, $eV >
\hbar\omega_0,$  the heating process makes a linear contribution to conductance and to the
relaxation rate:
\begin{equation}
G(V) =  e\alpha \left(1 + 2\nu_0{eV\over \hbar \omega_0}\right),  \gamma = {\hbar \alpha \over
m\lambda^2}\left(1 + 2\nu_0{eV\over \hbar \omega_0}\right).
\end{equation}

We consider here two models of an oscillator having different masses. For the first model, the shuttle,
the mass is $10^{-20}g$, and this model describes the molecules embedded in an elastic medium between the leads. The molecule is taken to be electrically connected to one lead, whereas
a connection to the other lead is realized by a tunnel junction.

The second model is associated with the cantilever having a mass  $10^{-12}g$ which is placed in the
neighborhood of the electrical contact; in this case variations of the cantilever position modulate
the tunneling matrix element of the junction.
 The oscillator frequencies are the same for both cases and are equal to $1 GHz, \hbar
 \omega_0 = 10^{-18} erg.$
The tunneling lengths are also the same both for the shuttle and the cantilever and are equal to
$10^{-8}cm$. As a result, the parameter $\nu_0$ is much less than one ($\sim 5\cdot 10^{-7}$) for the
cantilever and is of the order of one ($\sim 0.5$) for the shuttle.

To characterize the nonequilibrium fluctuations of the systems, we plot the bias and temperature dependencies of the
parameter
$$\tilde{\nu}_c= {\nu_c \over \nu_0\coth(\hbar\omega_0/2T)} = {2m\omega_0\langle\tilde{x}^2\rangle\over
\hbar\coth(\hbar\omega_0/2T)},$$ i.e. the ratio of the dispersion fluctuations of the oscillator
coupled to the tunnel junction to that of the uncoupled oscillator at temperature $T$.

The bias dependencies at various temperatures are presented in Fig.1(a) for the cantilever and in
Fig.1(b) for the shuttle. One can see that at extremely low temperature there is the change of the
curve shape at the critical voltage corresponding to the characteristic frequency of oscillator. For
the cantilever, there is not any heating below this bias and there is linear growth of fluctuations
thereafter, as indicated by the analytical results above. As temperature increases the curve becomes smoother and heating decreases, and,
finally, there is almost no heating at $T=0.1K$. For the shuttle case, there is some heating even
below the critical voltage and, moreover, the equilibrium value (at $V=0$) of $\tilde{\nu}_c$ is not one. The reason for this is that at $\nu_0\sim 1 $, vacuum fluctuations in the leads (back and forth tunneling events) give rise to oscillator anharmonicity and the steady-state fluctuation level of
the oscillator nonlinearly coupled to the electron reservoirs is different from that of the uncoupled
oscillator.

The temperature dependencies are exhibited in Figs.2, 3(a,b) and 4(a,b) for the cantilever, and similar
features have been found for the shuttle. The bias voltages are chosen to be below (Fig.2), above
(Fig.3(a,b)) and far above (Fig.4(a,b)) that of the excitation threshold  $\hbar\omega_0/e = 0.625 \times
10^{-6} V$, respectively. The "b" parts of the figures magnify the low temperature behavior. It is evident from these figures that as temperature
increases, $\tilde{\nu}_c$ approaches one for any bias voltage. However, for low bias there is a
peak at low temperature which becomes more pronounced below the critical bias voltage. This peak
occurs when temperature is such that $T\sim \omega_0 -eV$; the peak is absent at high voltage. This phenomenon occurs for both the cantilever and the shuttle.

\section{Summary}

In summary, we have applied the theory of open quantum system \cite{Efremov1,Efremov2,Smirnov1} to the nanoelectromechanical system
consisting of an oscillator coupled to a tunnel junction. We have obtained explicit expressions
for the nonequilibrium fluctuation level of the oscillator (Eqs.(28),(48)) as well as for the tunnel current
through the structure (Eq.(45)) and for the decoherence rate of the oscillator (Eqs.(25),(46)). The bias and
temperature dependencies of the oscillator fluctuation level have been determined. Considering two specific models with
different oscillator masses, we have shown that, for small mass, the level of mechanical vacuum fluctuations in this system differs significantly from that of a linear harmonic oscillator with the
same frequency $\omega_0$ because of an anharmonicity induced by the weak nonlinear oscillator-current
coupling. This effect takes place even at zero bias voltage, but vanishes at higher temperatures. We have
also found that for voltages below the excitation threshold, $\hbar\omega_0/e$, the relative level of oscillator fluctuations (normalized to the equilibrium value) peaks at
temperatures of the order of the difference between the characteristic frequency of the oscillator and the applied bias.

\begin{center}
{\bf Acknowledgement}
\end{center}

We are thankful to Dima Mozyrsky for valuable discussions which brought our interest to this problem.
L.G.M. and N.J.M.H. gratefully acknowledges support from the Department of Defense, DAAD 19-01-1-0592.

\begin{figure}[tbp]
\caption{Voltage dependence of the nonequilibrium fluctuations of oscillator position (normalized to those of the uncoupled oscillator) at various temperatures; (a) for oscillator mass $m=10^{-12} g $,
(b) for oscillator mass $m=10^{-20} g $.} 
\label{fig1}
\end{figure}

\begin{figure}[tbp]
\caption{Temperature dependence of the nonequilibrium fluctuations of oscillator position (normalized to those of the uncoupled oscillator) for bias voltage $10^{-7}V$.}
\label{fig2}
\end{figure}

\begin{figure}[tbp]
\caption{(a) Temperature dependence of the nonequilibrium fluctuations of oscillator position (normalized to those of the uncoupled oscillator) for bias voltage $10^{-6}V$; (b) Magnified low temperature part of this dependence.}
\label{fig3}
\end{figure}

\begin{figure}[tbp]
\caption{(a) Temperature dependence of the nonequilibrium fluctuations of oscillator position (normalized to those of the uncoupled oscillator) for bias voltage $10^{-4}V$; (b) Magnified low temperature part of this dependence.}
\label{fig4}
\end{figure}

\end{document}